\documentclass[aps,prb,twocolumn,groupedaddress,showpacs,floatfix]{revtex4}
\usepackage{amsmath}
\usepackage{graphicx}
\usepackage{bm}

\begin{document}
\title{Review of the recent x-ray and neutron powder diffraction studies on lead zirconate titanate}


\author{J. Frantti}
\affiliation{Laboratory of Physics, Helsinki University of Technology, P.O. Box 4100, FIN-02015 HUT, Finland}
\email{jfr@fyslab.hut.fi}

\date{\today}

\begin{abstract}
The issues related to the structure refinement of Pb(Zr$_x$Ti$_{1-x}$)O$_3$ (PZT) solid solutions 
are discussed. Particular attention is paid on the modelling of the co-existing phases in 
the vicinity of the morphotropic phase boundary (MPB), where local symmetry is often 
significantly lower than the average symmetry. According to recent studies, two-phase co-existence 
in the vicinity of the MPB is a thermodynamical necessity. Significantly different structural 
models for PZT with $x$ in the vicinity of the MPB have recently been published. Structural 
models, based on x-ray, neutron and electron diffraction studies, are reviewed and two 
essentially different approaches were identified: (i) a method where space group symmetry 
was decreased untill the features due to the local distortions were 'modelled' and (ii) a 
method where the highest space group symmetry compatible with the powder diffraction data 
was used together with a model for a local disorder. Related to the method (ii), the essential 
features of a model taking the $hkl$-dependent line broadening into account are summarized. 
The importance of the cationic disorder is demonstrated by computing the variation in Madelung 
energies due to the variation in structural parameters for PZT using experimental data.
The underlying theme is to consider how local regions, with symmetry lower than the average 
symmetry, should be modelled. Method (i) often introduces unjustified structural parameters 
and should be avoided. By studying the temperature and composition dependence of the structural 
parameters, including line shape parameters and phase fractions in the case of the two phase samples, 
one can further test the proposed models. The significance of the local structure for electrical properties 
(switchable polarization and electrical conductivity) is discussed by considering the local Pb ion 
displacements and their dependence on the neighbouring $B$-cation configuration. The connection between 
the structure and piezoelectric properties, with emphasis on the co-existence of the rhombohedral and 
monoclinic phases, is discussed in light of the recent first-principles computational studies. Also 
problems connected to sample preparation and data collection are pointed out. 
\end{abstract}


\maketitle

\section{Introduction}

Lead zirconate titanate [Pb(Zr$_x$Ti$_{1-x}$)O$_3$, (PZT)] solid solutions are 
among the most widely used ferroelectric ceramics.  The importance of PZT ceramics with 
morphotropic phase boundary (MPB) compositions is due to the exceptionally good 
piezoelectric properties they exhibit \cite{Jaffe}, 
which in turn has motivated numerous studies seeking physical mechanisms responsible for this 
extraordinary behaviour (for a summary of the various studies see ref. \onlinecite{Glazer1}). 
Crystal structure studies 
and the changes in atomic positions versus composition, temperature, pressure or electric 
field have a central role for the understanding of these materials: nuclei positions, 
together with the electron density, determine the electric polarization vector on which the 
numerous applications are based on. It is essential to understand how the polarization vector 
changes as a function of electric field, composition, temperature or pressure. Qualitatively, the average structural 
tendencies versus composition, temperature or pressure of ionic compounds, 
such as PZT, can be understood by considering the Madelung energies which decreases with decreasing 
crystal volume until the shortrange interactions balance the crystal volume to its equilibrium value. 
For ionic crystals the contribution of Madelung energy to the total energy is dominant and implies 
that crystal volume tends to be minimized. By tilting the oxygen octahedra the crystal 
can partially compensate for the effect of a larger average $B$-cation (Zr and Ti) size which increases with 
increasing Zr content $x$. This is seen from the behaviour of Zr rich samples: once crystal adopts $R3c$ symmetry, 
instead of $R3m$ symmetry, the volume per primitive cell is smaller. This is consistent with the notion that 
in the case it would cost a lot of energy to contract the oxygen octahedra (due to the shortrange interactions 
which cause the mere compression of octahedra to be energetically unfavourable) overall energy gain can still 
be obtained by tilting the oxygen octahedra. 
This can neatly be expressed through the equation $V_A/V_B\approx 6\cos^2\langle \omega \rangle -1$, where $V_A$ and $V_B$ are 
the polyhedral volumes for $A$ and $B$ cations and $\langle \omega \rangle$ is the mean octahedral tilt angle \cite{Thomas}. 
This expression means that when there is no chance for oxygen octahedra to decrease its volume  
the crystal can still be compressed by decreasing the volume of the cuboctahedra surrounding the $A$ cation. 
Typically the changes in octahedral tilts occur consistently versus composition\cite{Corker} or temperature\cite{FranttiPRB} 
and thus particular care is necessary before anomalous octahedral tilt systems are considered. Secondly, in the case of PZT 
(and other Pb containing perovskite ferroelectrics) the crystals suffer from numerous defects. These defects do have a 
significant role for ferroelectricity. An example is provided by the Pb ions which are significantly 
displaced from their ideal sites and which respond to the external electric field or pressure by adjusting their position 
with a corresponding changes in polarization vector \cite{NohedaA,NohedaB}. As was discussed in ref. \onlinecite{Bell}, it is 
not clear if Pb displacements are dominantly dynamic or static. 
Static displacements have commonly been modelled by displacing Pb ions 
towards $\langle 110 \rangle$ directions and using isotropic atomic displacement 
parameters (ADP), whereas the dynamic contribution was modelled by anisotropic ADP parameters. In some cases, one can try to 
distinguish the static and dynamic contributions by collecting structural data as a function of temperature\cite{ITD}. In PZT 
and PbTiO$_3$ crystals the contribution of anharmonic terms to the potential is significant, as was demonstrated by Raman 
scattering studies in the case of PbTiO$_3$ \cite{Foster,FranttiPRBPT} and PZT \cite{FranttiPRBPZT}, and obscures the separation 
of static and dynamic contributions.  
Similar situation occurs in many Pb and Bi containing perovskites and is often believed 
to be an important factor allowing a co-existence of ferromagnetic and ferroelectric ordering in multiferroic perovskites \cite{Hill}. 
Related consideration holds for Zr and Ti, whose positions are disordered in two ways: 
(i) the probability that the $B$-cation site is occupied by Zr (Ti) is $x$ ($1-x$) and (ii) the fractional coordinates of Zr and Ti are 
different. The confirmation of the assumption (i) is not straightforward, but various powder diffraction and Raman scattering studies 
show that it is a good approximation for most cases. By high resolution powder diffraction instruments it is possible to study these 
peculiar features by extracting information from the line shapes, in addition to the determination of the average structure. However, it 
became apparent that a rather complex lineshape is necessary for the modelling of the diffraction 
peak profiles, for examples see refs. \onlinecite{Stephens} and \onlinecite{Leineweber}.
The structure of the PZT ceramics with composition in the vicinity of the MPB are notoriously tricky 
to model. In addition to the aforementioned structural distortions there are two other reasons obscuring the modelling work: 
(i) in practice, no single phase samples are obtained, instead one observes two perovskite phases and 
(ii) diffraction peak profiles are rather complex. Both factors are connected to the fact that this 
'boundary' separates tetragonal and rhombohedral phases which do not possess a group-subgroup 
relationship. The experimental observation that monoclinic $Cm$ phase separates tetragonal 
(space group $P4mm$) and rhombohedral (at room temperature space group is $R3m$) phases clarified the 
situation: $Cm$ is a common subgroup of $P4mm$ and $R3m$ phases\cite{NohedaA,NohedaB}. Correspondingly, 
instead of morphotropic phase boundary one occasionally refers to morphotropic phase\cite{NohedaA,NohedaB}.
Thus, it is not a surprise that the structural disorder is particularly pronounced in the vicinity of MPB. 
This is also partially due to fact that typically the structure and composition of the grain boundaries and 
bulk are different. This is revealed from the Bragg reflection line shapes, which are almost 'Gaussian' like 
at room temperature \cite{NohedaB,FranttiJJAP} in the case of the less abundant phase. 
Even though it is still an open question whether $Cm$ is stable phase in PZT system it provides an explanation 
to the peak value of piezoelectric constant via the polarization rotation theory: first-principles computations 
demonstrated that the polarization rotation provides the lowest free energy path along which there is a large 
coupling between the polarization and the electric field\cite{Fu}, consistently with the experimentally observed 
large electromechanical response. Thermodynamical consideration of the polarization rotation within monoclinic phase 
and its impact on piezoelectric properties is given in ref. \onlinecite{Bell}. This study addresses the question if it is necessary to invoke 
the monoclinic phase to explain the observed piezoelectric properties.

Recently, a model which used two monoclinic phases (space group symmetries $Cc$ and $Cm$) was claimed to be the correct 
low-temperature structure of PZT with $x\approx 0.52$ \cite{Ranjan}, see also note \onlinecite{Note2} and refs. \onlinecite{Ranjan3} 
and \onlinecite{Hatch}. This report stated that $R3c$ symmetry is not a correct choice for PZT with $x\approx 0.52$ \cite{Ranjan}. 
The fact that superlattice reflections allow the distinction of $R3c$ and $Cc$ symmetries in favour of the former was overlooked 
in this study. Nevertheless, $Cc$ symmetry predicts reflections well isolated from those of the $R3c$ symmetry. To summarize the 
models proposed in refs. \onlinecite{Ranjan} and \onlinecite{Ranjan3}, we note that the phase which traditionally has been assigned 
to the rhombohedral symmetry\cite{Jaffe}, was assigned to $Cm$ symmetry and space group symmetry $P4mm$ (at room temperature) or 
$Cc$ (at low temperature) was assigned to the phase which was modelled by $Cm$ symmetry in refs. \onlinecite{FranttiPRB,NohedaA} 
and \onlinecite{NohedaB}. Thus, significant changes to the well known 
PZT phase diagram\cite{Jaffe} were proposed. To understand the problems related to such a model (abbreviated as $Cc+Cm$) and the 
reason why the use of space group $Cc$ was previously rejected \cite{FranttiPRB} it is necessary to summarize the crucial role of 
the local disorder resulting in $hkl$-dependent line broadening \cite{Note0}. Erroneous space group symmetry assignments likely 
result in once this line broadening is compensated for by reducing the space group symmetry\cite{Note1}.
The $hkl$-dependent line broadening is \emph{an inherent} property of PZT powders due to the local distortions, and is already seen in 
tetragonal Ti rich compositions and even at high temperature cubic phase\cite{FranttiJPCM}. 
Namely, the Bragg reflection widths were anisotropic already at the cubic phase, although the linewidths 
were an order of magnitude smaller than in the case of the low-temperature phases. 
It is well known that broad peaks are observed in Raman spectra collected at high temperature cubic 
phase, even though no first order Raman scattering is allowed for the ideal symmetry. For example, see
 ref. \onlinecite{Ikeuchi}, where the modes appearing in cubic phase were assigned to the first order scattering, 
activated by the local disorder, by studying the temperature dependence of the intensities. The 
same behaviour was observed in a closely related Pb(Hf$_x$Ti$_{1-x}$)O$_3$ (PHT) system with $0.10\leq x \leq 0.40$, 
where the peak widths of the $00l$ reflections were \emph{twice} as large as 
the widths of $h00$ reflections\cite{FranttiSub}. It must be emphasized that 
no signs of symmetry lowering from the $P4mm$ symmetry was observed in this 
high-resolution neutron powder diffraction study. In contrast, light scattering experiments 
revealed that both in the case of tetragonal PZT and PHT samples deviations from the 
average symmetry were observed \cite{FranttiPRB2,FranttiJJAP2,FranttiBoston}. Namely, 
the number of Raman active modes was larger than the symmetry found by diffraction 
techniques would allow. Once the diffraction patterns were examined it became apparent 
that there was no justification to use lower space group symmetry. A model based on 
lower space group symmetry would simply result in physically meaningless structural 
parameters. In this case, the essential point was that even if one insist to invoke 
lower space group symmetry in order to explain the observed Raman spectra the changes 
in bond lengths should be so large that also high resolution neutron powder diffraction 
patterns should reveal it. Although one can decrease the average symmetry to decrease the 
residuals, a more realistic model is obtained by assuming that (i) the \emph{average symmetry} 
over a length scale of a few hundred nanometers is $P4mm$, and (ii) deviations from 
this symmetry occur in a local scale (of a few unit cells). The essential point 
here is that these deviations are not periodical, at least not in a scale probed by x-ray or neutron diffraction. 
However, ion displacements (particularly those of Pb ions) are likely correlated in a scale of a few unit cells\cite{Glazer1}. 
It is also worth mentioning that usually there are several line broadening mechanisms. Especially powders consisted of fine 
crystallites, typically obtained through wet chemical methods allowing a preparation of perovskite phase at rather moderate 
temperatures, show large broadening due to the small crystallite size, in addition to the anisotropic line broadening \cite{RossettiXRD}. 
Williamson-Hall plot was constructed to separate crystallite and anisotropic strain broadening effects\cite{RossettiXRD} in tetragonal PZT. 
This study revealed that the lattice strain is greater along the $[001]$ directions than along the $[100]$ directions. On the other hand, 
samples prepared through solid state reaction often have an order of magnitude larger crystallite size. In the case of PZT with compositions 
in the vicinity of the MPB the line widths are strongly increasing with decreasing temperature\cite{FranttiJPCM}, demonstrating 
that the crystallite size and shape alone cannot account for the observed line broadening behaviour. Similar, although weaker, 
phenomenon occurs in the case of tetragonal PZT and PHT samples. 

Transmission electron microscopy (TEM) and electron diffraction (ED) studies revealed extra reflections in rhombohedral 
PZTs \cite{Viehland}. The origin of these reflections was more recently studied by TEM and ED together with neutron powder 
diffraction (NPD) techniques \cite{Glazer}. In this study superlattice reflections of the type $R_1$, $R_2$, $M_1$ and $M_2$ 
(see Table \ref{ED}) were observed, consistently with ref. \onlinecite{Viehland}. Among these reflections only 
$R_2$ is consistent with $R3c$ symmetry \cite{Glazer}. We summarize the observations and interpretations given in ref. 
\onlinecite{Glazer} as follows: (i) $M$-type reflections were accompanied by satellites, (ii) different areas in the 
same grain resulted in changes in the relative intensities of the superlattice reflections so that in some cases the extra 
spots completely disappeared, (iii) by collecting data on samples prepared from single crystals and ceramics it was 
found that the satellites around $M$ points are characteristic feature of only ceramics, (iv) by studying the temperature 
dependence of the extra reflections they could be assigned to the ferroelectric state (no extra reflections were observed 
at cubic phase), (v) octahedral tilts and distortions alone are insufficient to explain $R_1$ and $M$ superlattice 
reflections \emph{with the intensities observed} (it was demonstrated that very weak superlattice reflections could be obtained 
using unrealistically large oxygen octahedra tilts and distortions), and (vi) NPD experiments revealed \emph{only} $R_2$ 
reflections, consistently with an average $R3c$ symmetry. To explain these observations models based on locally ordered 
regions presenting antiparallel cation displacements were proposed (see also ref. \onlinecite{Corker} where structural models based 
on NPD experiments are given) and compared with experiments. Finite size effect was proposed as an underlying reason for the 
phenomena observed by TEM and ED techniques. This is an important point since it suggests that \emph{these puzzling phenomena, 
observed by TEM and ED techniques, do not occur in the case of bulk ceramics}. However, the tendency of Pb ions to form four 
short bonds with oxygen is a feature observed through NPD studies \cite{Corker}. This tendency is also seen from the models 
constructed  to explain the observed ED patterns \cite{Glazer}. It is also worth to note that similar conclusions were 
obtained through Monte Carlo simulations: $M$-like superlattice reflections are not present in 
pure bulk crystals but might be locally induced by surfaces over some range of temperature\cite{Leung}.
\begin{table}
\begin{center}
\caption{\label{ED}. Classification of superlattice reflections commonly found in the rhombohedral phases of PZT by TEM/ED studies. 
Tilts refer to Glazer notation\cite{GlazerNotation}. Table is adapted from ref. \onlinecite{Glazer}.} 
\begin{tabular}{l l l l}
\hline
Type        & Reflection             & Conditions           & Tilt \\
\hline 
R$_1$       & $\frac{1}{2}\{hkl\}_p$ & $h=k=l$              & None  \\
R$_2$       & $\frac{1}{2}\{hkl\}_p$ & $h \neq k \neq l$    & $a^-$ \\
M$_1$       & $\frac{1}{2}\{0kl\}_p$ & $k=l$                & None  \\
M$_2$       & $\frac{1}{2}\{0kl\}_p$ & $k \neq l$           & $a^+$ \\
\hline
\end{tabular}
\end{center}
\end{table}

After a reasonable model for local distortions or locally ordered regions is identified a selection of an appropriate instrument 
for the (preferably neutron) powder diffraction experiments to further test the structural models is essential. 
For example, the $Cm$ phase was identified by studying the peak split of the pseudo-cubic $110$ reflections 
\cite{NohedaA,NohedaB} through high resolution synchrotron experiments. This peak split was beyond the 
resolution of the instrument used in ref. \onlinecite{Ranjan} and there is no way to recover the lost information. 
Thus, new refinements do not compensate for the problems connected to the low-resolution data collected with 
insufficient counting time. The role of the spatial composition variation (which in practice cannot be 
eliminated) in the vicinity of the phase boundary is an old and still a valid explanation for the two phase 
'co-existences' and also largely explains the $hkl$-dependent line broadening. A qualitative model which was 
constructed by studying changes in structures and their mutual phase fractions versus temperature and $x$ was 
proposed in refs. \onlinecite{FranttiPRB} and \onlinecite{FranttiJPCM}. These aspects are reviewed below. 
In section \ref{MadelungEnergy} the variance in Madelung energies, corresponding  
to the variance in experimentally determined structural parameters, is given for PZT. The idea of this 
consideration is not to give a qualitative treatment for the total energy of disordered PZT system but to 
demonstrate the crucial role of disorder by showing that rather small variations in structural parameters 
correspond to large variation in Madelung energy. 
Various Pb-ion displacement configurations observed in different phases of PZTs are addressed in section \ref{Pb}. 
This section reviews the physical origin of Pb-ion displacement and its effect on modelling and electrical properties. 
In section \ref{SCV} 
we discuss the role of the spatial composition variation on the PZTs with $x$ in the vicinity of the MPB, as it 
is closely connected to the two phase co-existence. The main goal of the section \ref{SCV} is to provide an explanation 
for the different type of domains by taking into account the recent thermodynamical considerations and to show that 
these ideas are consistent with numerous experimental observations. Anisotropic line broadening is discussed in section 
\ref{ALB}. Section \ref{TPM} reviews different two-phase models proposed for PZT in the vicinity of MPB. Section \ref{BVS} 
summarizes the results obtained through bond-valence computations and recent computational studies.

\section{\label{MadelungEnergy}Madelung energies for PZT system}
The Madelung energy was computed by applying the Ewald method \cite{Born,Cowley}. The Madelung energy is 
given by the equation $E_M = \sum_{kk'}z_kz_k'\alpha_{kk'}/(\epsilon_0 r)$, where the summation runs over all ions $k$ and 
$k'$ (with charges $z_k$ and $z_{k'}$, respectively) in a primitive cell (assumed to be charge neutral) and  $\epsilon_0$ is 
the permittivity of free space. Charges were assumed to be integer multiples of the electric charge $e$. 
Following refs. 
\onlinecite{Born} and \onlinecite{Cowley} the coefficients $\alpha_{kk'}$ are written as 
\begin{eqnarray}
\label{Madelung}
\lefteqn{\alpha_{kk'}=\frac{1}{2}c{\sum_{l'}}'H(c|\mathbf{x}_{l'k'}-\mathbf{x}_{0k}|/r)-\delta_{kk'}c/\sqrt{\pi}+}
\nonumber\\
& &
+\pi/(2c^2s){\sum_{hkl}}'G(\pi^2|\mathbf{b}(hkl)|^2r^2/c^2)\exp[2\pi i \mathbf{b}(hkl)\cdot
\nonumber\\
& & (\mathbf{x}_{0k'}-\mathbf{x}_{0k})r^2/c^2],
\end{eqnarray}
where $H(x)=(1/x)$ $\mathrm{erfc}(x)$ ($\mathrm{erfc}$ is the complementary error function), $G(x)=(1/x)\exp(x)$, 
$\delta_{kk'}$ is the Kronecker delta function, $\mathbf{x}_{lk}$ is the position of the ion $k$ at the primitive 
cell $l$, $\mathbf{b}(hkl)$ is a primitive reciprocal lattice vector (indexed by integers $h,k,l$ and defined without 
the prefactor $2\pi$) and $v$ is the volume of the primitive cell, $v=sr^3$. Primes above the summation signs 
indicate that in the first (real space) sum the term $\mathbf{x}_{l'k'}=\mathbf{x}_{0k}$ is omitted, and in the 
case of the second (reciprocal lattice) sum the term $\mathbf{b}(hkl)=0$ is omitted. Number $c$ controls the 
convergence of the two sums, and was selected so that both the real and reciprocal lattice sums were of the same 
order of magnitude. The coefficients $\alpha_{kk'}$ do not depend on the value of $c$ ($d\alpha_{kk'}/dc$ 
vanishes for all $c$). However, by selecting too small or too large value for $c$ results in a situation where 
either the reciprocal or real lattice sum vanishes (within the numerical accuracy of the computational routines), 
leading to erroneous values for $\alpha_{kk'}$. We fixed $c$ at $\pi$. 

Crystal structure data for PZT ceramics were adapted from refs. \onlinecite{FranttiPRB,FranttiJPCM} and \onlinecite{Haines} 
(high pressure data). To simplify the computational task, we assumed that the 
fractional $z$ coordinates for both $B$\mbox{  }cations are the same (though they differed in the case of the 
tetragonal perovskites, see Table VI in ref. \onlinecite{FranttiPRB}). Thus, for this case the $z$ coordinate values 
for the $B$\mbox{ } cations were computed using an equation $z(B)=x\times z(\textrm{Zr})+(1-x)\times z(\mathrm{Ti})$ 
(this approximation was used for $x=0.20,0.30$ and $0.40$ data). Similar approximation was also done for the 
fractional $x$ coordinates of the $x=0.52$ PZT sample. In addition, we assumed that Pb ions were fixed at 
origin. During the Rietveld refinement they were allowed to shift towards the $\langle 110 \rangle$ directions. 
Ionic charges were fixed at their nominal values ($+2$ for Pb, $+4$ for the $B$ cations, and $-2$\mbox{  }for O). 
An error estimation for the Madelung energy was carried out using the standard deviation values of the structural parameters. 
The error estimates of the structural parameters were assumed to be independent. It is worth to note that the 
error estimates are specific to the structural model. For instance, for the cubic symmetry there is only one 
structural parameter, the $a$ axis value, while there are 11 structural parameters for $Cm$ phase. This resulted 
in significantly larger error estimates for $Cm$ phase.

\begin{figure*}
\begin{center}
\includegraphics[width=14cm,angle=0]{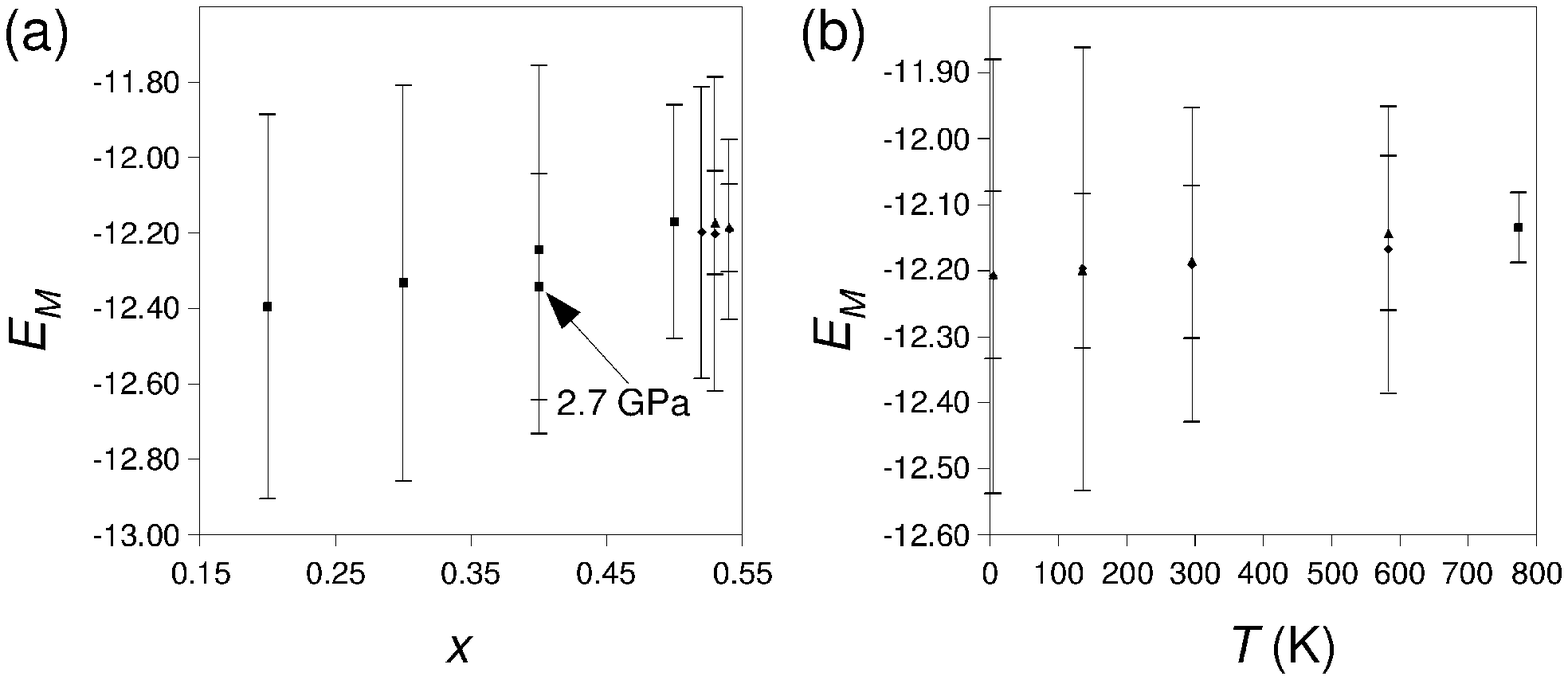}
\caption{\label{MadelungPZT} The Madelung energies (in units of $\epsilon_0 \mathrm{\AA}/e^2$) 
of PZT crystals per formula unit (a) versus composition $x$ at room temperature, (b) versus temperature 
in the case of the $x=0.54$ sample. $E_M$ values correspond to $P4mm$ phase for $x\leq 0.50$ and, in the 
case of the $x=0.52,0.53$ and $0.54$ samples, values for $Cm$ (squares) and $R3c$ (triangles) phases are 
given. The 773 K value corresponds to $Pm\bar{3}m$ phase. For $x=0.40$\mbox{  }sample, the structural data for 2.7 GPa 
value was adapted from ref. \protect\onlinecite{Haines}. In the case of the two phase samples, the larger 
error bars correspond to $Cm$ phase, see text.}
\end{center}
\end{figure*}

Figure \ref{MadelungPZT} shows the room temperature Madelung energies for PZT ceramics versus 
$x$ and, for the $x=0.54$ sample, versus temperature. The Madelung energy increases almost linearly with 
increasing $x$ (when $x\leq 0.50$), as is seen from Fig. \ref{MadelungPZT} (a). 
This is related to the fact that the average bond lengths increase with increasing $x$. 
However, octahedral tilt ($R3c$ phase) and collective Pb ion shifts ($Cm$ phase) correspond to a small energy gain. 
Both are features which were neglected in the present treatment of tetragonal PZT.
It was interesting 
to note that, in the vicinity of the MPB, $P4mm$ ($x=0.50$), $Cm$ and $R3c$ phases ($x=0.52$) had almost 
the same Madelung energies: $Cm$ phase had the lowest energy, while the energy of $R3c$ phase was almost the 
same, and approached the energy of $Cm$ phase with increasing $x$ (Fig. \ref{MadelungPZT} (a)). 
This is consistent with the observed phase transition sequence against $x$. As the difference between 
Madelung energies of $Cm$\mbox{  }and $R3c$ phases in the vicinity of MPB is small, it is not a 
surprise that a preparation of single phase samples within this composition range is very difficult. Also 
the evolution of Madelung energies versus temperature indicates that the difference between the Madelung energies 
of $Cm$ and $R3c$\mbox{ } phases is small up to room temperature, see Fig. \ref{MadelungPZT} (b). 
The Madelung energy of $R3c$ phase at 583 K should be treated with caution, since $R3c$ phase 
fraction at this temperature was very small (correspondingly, the structural data were not 
very accurate). It is clear from Fig. \ref{MadelungPZT} (b) that the Madelung energy 
solely does not explain the observed temperature dependent phase fractions seen in PZT with 
$x=0.54$ sample \cite{FranttiJPCM}: $Cm$ and $R3c$ phases had almost the same Madelung 
energies, and it was only at around 583 K that the difference became significant.
Although the given treatment is rather crude, it points out that both local cation shifts and octahedral tilts 
are mechanisms to be considered. The first one is significant through the whole composition range whereas the latter cannot be neglected at higher Zr concentrations. 

\section{\label{Pb}Local disorder: Pb ion displacements versus $x$}

\begin{figure*}
\begin{center}
\includegraphics[width=14cm]{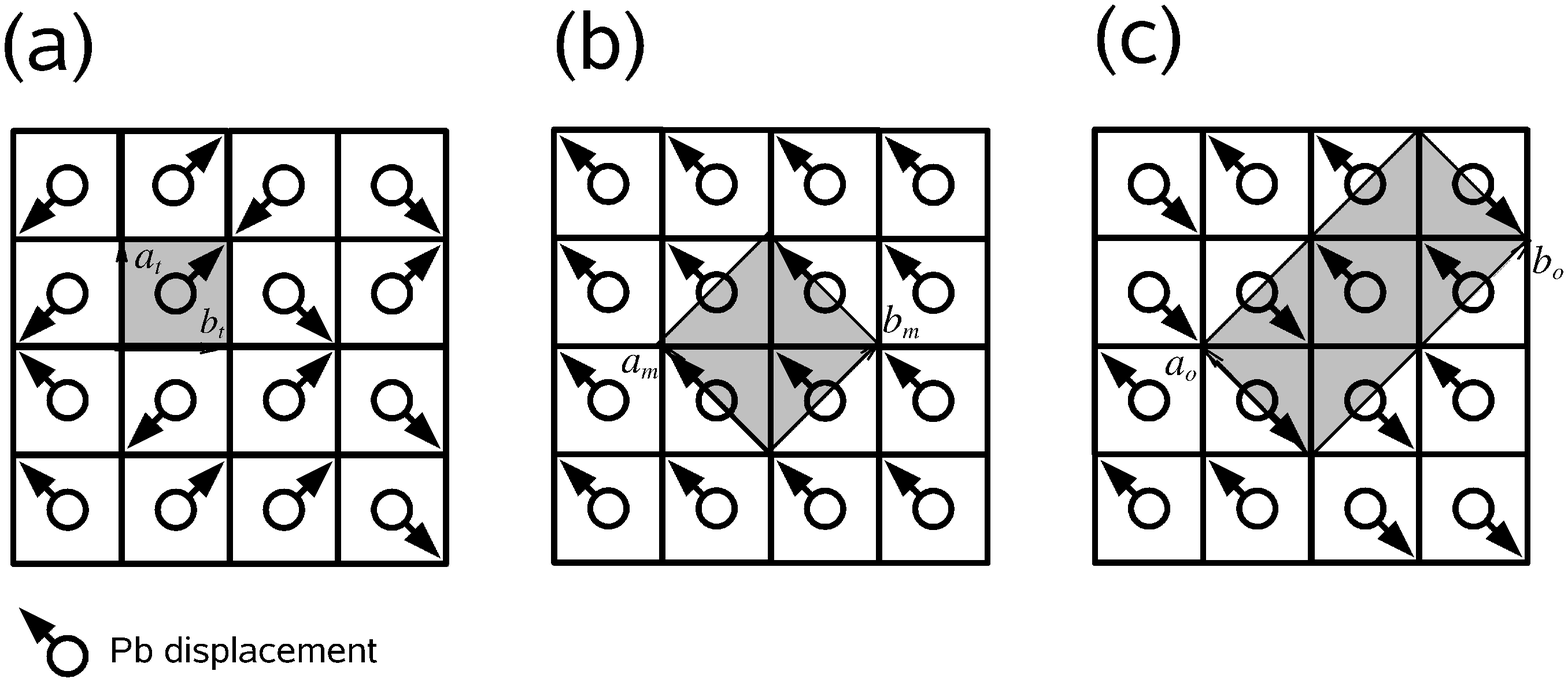}
\caption{\label{PbDisplacements} Pb-ion displacements in (a) tetragonal (space group $P4mm$), (b) monoclinic 
(space group $Cm$) and (c) orthorhombic (space group $Pbam$) phases. Shaded areas indicate the unit cells.}
\end{center}
\end{figure*}

The physical origin of the Pb-ion displacement in a local scale is often connected to 
the $6s^2$ lone electron pair ($L$) which makes displaced position energetically more favourable 
(actually one must also consider the $6p$ states, as discussed below). An insight to the important role of 
$L$ can be obtained by considering the formation of PbTiO$_3$ 
by alloying the litharge phase of PbO and TiO$_2$, as shown in ref. \onlinecite{LeBellac}. 
In the case of Pb$_{1-x}$(TiO)$_x$O solid solution a 
TiO$_2$ group is substitued for a PbO$L$ group and the extra oxygen fills the volume of 
$L$. Thus, due to $L$ Pb prefers to form four short 
Pb-O bonds in a pyramidal configuration, a tendency which is also seen in different 
phases of PZT: to fulfill this criteria Pb ions are 
displaced from their average positions (although the resulting PbO$_4$ pyramid is no 
more symmetric). As was discussed in ref. \onlinecite{Corker}, 
this occurs in the orthorhombic, tetragonal and rhombohedral phases of PZT. In other words 
this means that Pb ion is not located at the centre 
of the oxygen cuboctahedra, not even in the cubic phase. In ref. \onlinecite{Corker} local 
structures capable of producing the extra reflections 
of the type $\{\frac{1}{2},\frac{1}{2},0\}_p$ (Fig. 7) and $\{\frac{1}{2},\frac{1}{2},\frac{1}{2}\}_p$ 
(Fig. 8) seen in TEM and ED studies 
\emph{at approximately the correct intensities} are given. This picture gives a qualitative idea 
concerning the Pb ion displacements with respect 
to its nearest neighbours. Correspondingly, one must consider the whole oxygen and $B$-cation 
network in order to determine the overall Pb-displacement 
pattern. Table \ref{StructureSummary} summarizes the average structures of PZT system with 
typical structural parameters. Fig. \ref{PbDisplacements} 
shows the commonly assumed Pb-ion displacements for tetragonal PZT, PZT with 
$x$ in the vicinity of the MPB ($Cm$ symmetry) and PbZrO$_3$. Fig. \ref{PbDisplacements} 
demonstrates that there are numerous ways to displace Pb ions to form short Pb-O bonds: 
in PbZrO$_3$ Pb ions are displaced in such a way that an antiferroelectric ordering results 
in in $ab$ plane\cite{Corker2}.
\begin{turnpage}
\begin{table*}
\begin{center}
\caption{\label{StructureSummary} Average space group symmetries and representative values for structural parameters reported for 
PZT system. Note that sometimes the positions of Zr and Ti were refined independently. In the case of the rhombohedral phases the 
structural parameters refer to the hexagonal setting. Customarily $R3c$ symmetry has been used to describe the $R3m$ phase by 
setting $e=0$ (correspondingly the $c$ axis is halved). Thus only $R3c$ phase is specified. Lattice vectors and oxygen octahedra 
tilts are given in terms of 
pseudocubic lattice vectors. $x_{\mathrm{Zr}}$ refers to Zr content. Deviations from these symmetries are discussed in text.} 
\begin{tabular}{l l l l l l l l l l l l}
\hline
Space        & Lattice vectors             & Ion    & Multiplicity and  & $x$         & $y$      &  $z$ & Tilt        & Typical values for structural parameters       & $T$/K & $x_{\mathrm{Zr}}$ & Ref.                     \\
group  &  &        & Wyckoff letter &             &          &      &             &                                                &       &                   &                          \\
\hline 
$Pm\bar{3}m$ & $\mathbf{a}_c=\mathbf{a}_p$ &        &          &             &          &      & $a^0a^0a^0$ & $a_c=4.080$\AA                                 & 773   & 0.54              & \onlinecite{FranttiJPCM} \\
             &                             & Pb     & $1a$     & $0$         & $0$      & $0$  &             &                                                &       &                   &                          \\
             &                             & Zr/Ti  & $1b$     & $1/2$       & $1/2$    & $1/2$&             &                                                &       &                   &                          \\
             &                             & O      & $3c$     & $1/2$       & $1/2$    & $0$  &             &                                                &       &                       &                      \\
$P4mm$       & $\mathbf{a}_t=\mathbf{a}_p$ &        &          &             &          &      & $a^0a^0a^0$ & $a_t=3.978$\AA, $c_t=4.149$\AA                 & 295   & 0.30              & \onlinecite{FranttiPRB}  \\
             & $\mathbf{c}_t=\mathbf{c}_p$ & Pb     & $4d$     & $x$         & $y$      & $0$  &             & $x=y=0.015$                                    &       &            
         &                          \\
             &                             & Zr/Ti  & $1b$     & $1/2$       & $1/2$    & $z$  &             & $z_{\mathrm{Zr}}=0.560,z_{\mathrm{Ti}}=0.547$  &       &                       &                      \\
             &                             & O(1)   & $1b$     & $1/2$       & $1/2$    & $z$  &             & $z_{\mathrm{O(1)}}=0.101$                      &       &                       &                      \\
             &                             & O(2)   & $2c$     & $1/2$       & $0$      & $z$  &             & $z_{\mathrm{O(2)}}=0.617$                      &       &                       &                      \\
$Cm$ & $\mathbf{a}_m=\mathbf{a}_p-\mathbf{b}_p$ &   &          &             &          &      & $a^0a^0a^0$ & $a_m=5.722$\AA, $b_m=5.710$\AA, $c_m=4.137$\AA,& 20    & 0.52                  & \onlinecite{NohedaB} \\
     & $\mathbf{b}_m=\mathbf{a}_p+\mathbf{b}_p$ & Pb& $2a$     & $0$         & $0$      & $0$  &             & $\beta=90.498^{\circ}$                         &       &                       &                      \\
             & $\mathbf{c}_m=\mathbf{c}_p$ & Zr/Ti  & $2a$     & $x$         & $0$      & $z$  &             & $x_{\mathrm{Zr/Ti}}=0.477,z_{\mathrm{Zr/Ti}}=0.551$ &  &                       &                      \\
             &                             & O(1)   & $2a$     & $x$         & $0$      & $z$  &             & $x_{\mathrm{O(1)}}=0.449, z_{\mathrm{O(1)}}=0.099$  &  &                       &                      \\
             &                             & O(2)   & $4b$     & $x$         & $y$      & $z$  & & $x_{\mathrm{O(2)}}=0.212,y_{\mathrm{O(2)}}=0.257,z_{\mathrm{O2}}=0.627$ & & 
             &                      \\
$R3c$ & $\mathbf{a}_h=\mathbf{a}_p-\mathbf{b}_p$ &  &          &             &          &      & $a^-a^-a^-$ & $a_h=5.824$\AA, $c_h=14.372$\AA                & 295   & 0.80                  & \onlinecite{Corker}  \\
& $\mathbf{b}_h=\mathbf{b}_p-\mathbf{c}_p$ & Pb     & $6a$     & $0$         & $0$      & $s+1/4$ &          & $s=0.032$                                      &       &                       &                       \\
& $\mathbf{c}_h=\mathbf{a}_p+\mathbf{b}_p+\mathbf{c}_p$& Zr/Ti  & $6a$     & $0$         & $0$      & $t$     &          & $t_{\mathrm{Zr}}=0.014,t_{\mathrm{Ti}}=0.025$  &       &          
             &                      \\
             &                             & O      & $18b$    & $1/6-2e-2d$\mbox{ } & $1/3-4d$ & $1/12$  &  & $e=0.012$, $d=-0.003$                          &       &          
             &                      \\
$Pbam$       & $\mathbf{a}_o=\mathbf{a}_p-\mathbf{b}_p$ & &    &             &          &         &$a^-a^-c^0$ & $a_o=5.884$\AA, $b_o=11.787$\AA, $c_o=8.231$\AA& 100   & 1                     & \onlinecite{Corker2} \\
             & $\mathbf{b}_o=2(\mathbf{a}_p+\mathbf{b}_p)$ & Pb(1) & $4g$ & $x$ & $y$   & $0$     &          & $x_{\mathrm{Pb(1)}}=0.699,y_{\mathrm{Pb(1)}}=0.130$ &  &                       &                      \\
             & $\mathbf{c}_o=2\mathbf{c}_p$ & Pb(2) & $4h$     & $x$         & $y$      & $1/2$   &          & $x_{\mathrm{Pb(2)}}=0.707,y_{\mathrm{Pb(2)}}=0.123$ &  &                       &                      \\
             &                            & Zr      & $8i$     & $x$         & $y$      & $z$     &          & $x_{\mathrm{Zr}}=0.242,y_{\mathrm{Zr}}=0.124,z_{\mathrm{Zr}}=0.249$ & &        &                      \\
             &                            & O$_1$   & $4g$     & $x$         & $y$      & $0$     &          & $x_{\mathrm{O(1)}}=0.296,y_{\mathrm{O(1)}}=0.097$   &  &          
             &                      \\
             &                            & O$_2$   & $4h$     & $x$         & $y$      & $1/2$   &          & $x_{\mathrm{O(2)}}=0.278,y_{\mathrm{O(2)}}=0.156$   &  &          
             &                      \\
             &                            & O$_3$   & $8i$     & $x$         & $y$      & $z$     &          & $x_{\mathrm{O(3)}}=0.036,y_{\mathrm{O(3)}}=0.262,z_{\mathrm{O(3)}}=0.220$ & &  &                      \\
             &                            & O$_4$   & $4f$     & $0$         & $1/2$    & $z$     &          & $z_{\mathrm{O(4)}}=0.297$                           &  &          
             &                      \\
             &                            & O$_5$   & $4e$     & $0$         & $0$      & $z$     &          & $z_{\mathrm{O(5)}}=0.270$                           &  &          
             &                      \\ 
\hline
\end{tabular}
\end{center}
\end{table*}
\end{turnpage}
 Rough but illustrative picture for this can be obtained by considering the tolerance factor 
$t=(R_A+R_{\mathrm{O}})/(\sqrt{2}(R_B+R_{\mathrm{O}}))$ where $R_A$, $R_B$ and $R_{\mathrm{O}}$ are the ionic 
radii for the $A$- and $B$-cations and oxygen. In the case of PbZrO$_3$ $t<1$ implying that the Pb ions is not 
able to fill the cuboctahedra, whereas Zr fills octahedra so tightly that it takes a larger share from the total 
volume by tilting the octahedra. In this sense one can 'construct' PbZrO$_3$ by (i) considering the large Zr ions, 
which results in a $a^-a^-c^0$ tilt corresponding to distorted oxygen octahedra and then (ii) displacing Pb ions 
inside the cuboctahedra in such a way that four short, essentially covalent, bonds with oxygen (constituting a 
PbO$_4$ 'pyramid') are formed. Very similar situation occurs in a closely related PbHfO$_3$ which is isostructural 
and has the same space group symmetry as PbZrO$_3$\cite{Corker3}. For clarity it must be said that the given treatment 
is meant to be illustrative and not to give any causality relationships (in ref. \onlinecite{Corker3} the type of octahedral 
tilting and deformation was seen as a way to accommodate the nature of the Pb-O bonding). Now, the PbO$_4$ pyramids alternate 
in the $ab$ plane in such a way that antiferroelectric ordering is formed. Also the Zr ions experience similar shifts, although 
the absolute variation in Zr-O bonds (at 100 K the shortest and largest Zr-O bonds were 2.41 and 3.105 \AA) is not that large as 
in the case of Pb-O bonds (at 100 K the shortest and largest Zr-O bonds were 2.044 and 2.205 \AA). In contrast to Pb ions Zr ions 
exhibit anit-parallel $[001]$ shifts\cite{Corker2}. Also the behaviour of the Pb(1) and Pb(2) ions (specified in table \ref{StructureSummary}
versus temperature was found to be different. Namely, at 100 K the Pb(1) ion shift was significantly larger that of the Pb(2). 
Thus changes in Pb(1) position were larger than those of Pb(2) once the cubic phase was approached. It is interesting to note that 
in the case of Pb(1) three of the Pb-O bonds in PbO$_4$ pyramid were much shorter than the fourth one so that Pb(1) is almost threefold coordinated. 
In this context it is worth to mention that $Pbam$ and $R3m$ phases do not have group-subgroup relationship nor is common subgroup reported to exists 
between the two phases (in contrast to the case of MPB). Interestingly, there is a very narrow (with respect to temperature and $x$) region where 
PbZrO$_3$ and PZT with very small amount of Ti with $Pbam$ symmetry transforms into ferroelectric rhombohedral phase at around 500 K. However, 
the stability 
region of this phase is very sensitive to impurities \cite{Whatmore} and Pb ion positions are disordered \cite{Teslic}. Both phase transitions 
$Pbam \rightarrow R3m$ and $R3m \rightarrow Pm\bar{3}m$ are first order transitions (first necessarily by the given symmetry arguments)\cite{Whatmore}.
Since the Pb ion displacements from their ideal cubic sites are dominant in PbZrO$_3$ (so that the Pb ion are shifted with respect to the Zr and 
oxygen ions, which remain at their ideal sites) they essentially measure the sublattice polarization. Thus it is possible to simplify the structural 
model which in turn allows one to write down a simple free energy expansion\cite{Whatmore} for PbZrO$_3$.

In contrast, Pb-ions displacements average out in tetragonal PZT so that there is no net polarization in $ab$ plane. 
This is often modelled by assuming that the four $(xx0)$ sites, equivalent in the case of $P4mm$ symmetry, are statictically 
occupied by Pb ion (this is contrasted to the case where Pb ion is located at $1a$ site and anisotropic ADPs are used). The 
structural models used in Rietveld refinements take this into account by assuming that each site is occupied by a quarter of 
Pb ion. In the case of PZT the four sites are not equivalent, but to a first approximation, the occupation probability of 
each site does depend on the local $B$-cation configuration (how the eight nearest $B$ cation sites are occupied). Thus, the Pb-ion 
displacements are different in the case of Zr- and Ti-rich areas. $Cm$ symmetry allows that spontaneous polarization has a non-zero 
value in the $a$-axis direction. In the case of ideal symmetry, this corresponds to coherent Pb-displacements. 
However, due to the local distortions following spatial composition variation (and changes in bond lengths) also other Pb ion 
displacements (local disorder) must occur. Thus, there is a delicate interplay between $B$-cations, oxygen octahedra tilts (if any) 
and the corresponding Pb-ion displacements. It is not straightforward to say which displacement configuration 
is energetically the most favourable and is most reliably addressed through experiments. This is demonstrated by PbZrO$_3$ and PbHfO$_3$: 
due to the large space available for Pb one can construct various hypotetical PbO$_4$ configurations (different from the experimental 
ones, such as a one corresponding to a ferroelectric ordering) under the constraint that ions have their nominal valences. It is not 
obvious when a change to another structure would be favourable. These are aspects to be considered once new materials, such as multiferroics, 
are designed. 

PZT and Ba(Zr$_x$Ti$_{1-x}$)O$_3$ (BZT) systems provide interesting case study, since the changes in their electronic band structures versus $x$ are well 
documented. As was summarized in ref. \onlinecite{Warren}, there are two main differences between Pb perovskites and those containing closed-shell 
$A$-cations (such as Ba in the BZT system): 
(i) the band gap in PZT system is almost constant, being 3.45 eV in PbTiO$_3$ and 3.72 eV in PbZrO$_3$ whereas the band gaps in BaTiO$_3$ and BaZrO$_3$ 
are 3.0 and 5.0 eV, respectively and (ii) the presence of shallow Pb$^{3+}$ centers. To understand these observations, band structure computations 
and electron-paramagnetic-resonance (EPR) spectroscopy studies were carried out for PZT system, reported in refs. \onlinecite{Warren} and \onlinecite{Robertson} 
and summarized below. In BZT, the valence band edge consists of O $2p$ states and the conduction band consists of pure Ti/Zr $d$ states \cite{KingSmith}.
The band gap increases in BZT with $x$ because the Zr $4d$ states lie 2 eV above the Ti $3d$ states and Ba has little effect on the electronic structure 
because its states lie well away from the band gap. In PZT, the valence band has essentially the same width and character for all $x$, because this is 
determined by the Pb $s$-O $p$ interaction. Similarly to the case of BZT system, the Ti/Zr $d$ states increase rapidly in energy with $x$. 
The tight-binding computations for cubic PbTiO$_3$ and PbZrO$_3$ revealed that the conduction band minimum is composed of the Ti/Zr $d$ states only at 
low Zr compositions and switches to Pb $6p$ states with increasing $x$\cite{Warren,Robertson}. Shallow Pb$^{3+}$ centers corresponds to a local state 
slightly above the valence band maximum. EPR experiments revealed that the Pb$^{3+}$ center acquires more Pb $6p$ character with increasing $x$. This 
was interpreted to be due to the local off-center displacement of the Pb ion. Thus, by lowering the symmetry of the center the $p$ character is introduced 
into its wavefunction. The Pb$^{3+}$ hole trap binding energy was found to be between 0.14 and 
0.26 eV \cite{Robertson}. These finding were found to be consistent with Raman scattering studies of PZT, Nd-modified PbTiO$_3$ (PNT) and Nd-modified 
PZT (PNZT) bulk ceramics using ultraviolet (wavelength 363.79 nm which corresponds to 3.4 eV, close to the band gap energy) and visible light (wavelength 
514.532 nm which corresponds to 2.4 eV)\cite{FranttiUV}. In the case of uv light, several high-frequency modes above 1000 cm$^{-1}$ were observed. Most 
interestingly, the frequency of the mode at around 1170 cm$^{-1}$ in PZT ceramics showed anomalous behaviour at the well-known phase transitions, 
corresponding to the changes in Pb ion environment. The use of different systems (PZT, PNT and PNZT) allowed the assignment of these modes to an 
electronic processes in the Pb$^{3+}$ hole traps\cite{FranttiUV}, in contrast to Zr and Ti. Once this information is combined with the Pb ion displacements 
observed through NPD studies one gains very consistent picture of these systems. Now, these observations are important not only for the understanding 
factors affecting on switchable polarization in Pb-perovskites, but also provide a physical model explaining the activation energies $E_A$ of conductivity 
in PZT. Values between 0.15 and 0.18 eV were calculated for the $E_A$ assuming Poole-Frenkel mechanism for the conductivity \cite{LappalainenEC}. 
In many ways analogous situation appears in Bi-perovskites, such as in (BiScO$_3$)$_x$-(PbTiO$_3$)$_x$ (BS-PT) alloys\cite{Iniguez}. Namely, the large 
polarization and piezoelectric responses in BS-PT system were found to be due to the Bi/Pb $6p$-O $2p$ hybridization\cite{Iniguez}.

\section{\label{SCV}Spatial composition variation} 
The PZT system is frequently assumed to be a binary solid solution of PbTiO$_3$ and PbZrO$_3$ \emph{without 
solubility gaps}, as was summarized in ref. \onlinecite{Cao}. According to ref. \onlinecite{Cao}, once a solid solution is 
formed at high temperature (typically above 800$^{\circ}$C), the chemical composition cannot be changed at low 
temperatures, but the system can have temperature-induced \emph{diffusionless} structural phase transitions. 
The validity of this assumption has been questioned in refs. \onlinecite{Rane} and \onlinecite{Rossetti} and it was reported that PZT 
solutions exhibit a positive enthalpy of mixing that suggests \emph{a tendency towards immiscibility and phase 
decomposition}. Thus, it was proposed\cite{Rossetti} that miscibility gaps replace the MPB and the paraelectric 
to ferroelectric transition lines of the diffusionless phase diagram given in ref. \onlinecite{Jaffe} (which would be 
valid only if the cooling rate significantly exceeds the diffusion rate of Ti and Zr atoms). The analysis was 
restricted to temperatures above 420 K\cite{Rossetti}, although very similar conclusions should be valid at 
lower temperatures. Usually the solubility gap increases with decreasing temperature. In the present case the 
solubility gap and finite atomic diffusion rate would imply that system minimizes its free energy by segregating into two 
different phases possessing different compositions and crystal symmetries. Although it is probable that the equilibrium 
situation is not necessarily reached through standard sample preparation techniques, two-phase 
co-existence is apparent in the vicinity of the MPB. From the microscopic point of view this means that attention should 
be paid on how Zr and Ti ions are distributed over the $B$-cation site. This issue was recently addressed in ref. 
\onlinecite{Bell} through Monte Carlo type model where the mean of the distribution of cluster volume was analysed as a 
function of $x$. This study revealed that the sum of the mean cluster size of the Zr and Ti 
clusters is \emph{a minimum} at $x=0.50$, with the most likely minimum volume in which an equal number of Zr and Ti ions is 
found being 1 nm$^3$ (for clarity we point out that correspondingly there were many 1 nm$^3$ volumes with inequal number of 
Zr and Ti ions). Correspondingly, it was argued that the increase in coherence length of the nano-domains, proposed 
as the mechanism for the appearance of the macroscopic monoclinic phase close to $x=0.50$ in ref. \onlinecite{Glazer1}, may be 
a consequence of some ordering of the $B$-cations. According to the model given in ref. \onlinecite{Glazer1}, the local structures 
of the $R3m$ and $P4mm$ phases were considered to be monoclinic and $Cm$ phase was interpreted to correspond to those 
composition and temperature values for which the local regions had grown sufficiently that diffraction techniques see a 
distinct $Cm$ phase. For the forthcoming discussion, it is important to notice that the spatial composition variation 
is larger the smaller the spatial length scale is, as is also apparent from Fig. 13 in ref. \onlinecite{Bell}. Our next goal 
is to estimate the magnitude of the spatial composition variation in simplest terms and to relate it to the experimental 
observations on local and average structure.

Thus, we assume that the distribution of Zr and Ti is random. For discussion purposes, we further assume that the spatial 
composition can be divided into two parts: (i) differences between the average compositions of grains, 
consisted of domains and (ii) composition variation within a domain. 
To quantify the meaning of spatial composition variation within a domain we assume that the distribution of Zr and Ti obeys the 
binomial density function with probabilities $p=x$ and $q=1-p$ that the $B$-cation site is occupied by Zr and Ti 
ion, respectively (the solubility gap would further enhance the spatial composition variation). In principle, ordering effects can be taken 
into account by introducing conditional probabilities. We believe that the present approximation provides a microscopic explanation for the two 
phase co-existence observed in the vicinity of the MPB and is consistent with the idea that $Cm$ phase serves as a transitional 
phase between rhombohedral and tetragonal phases. To see this, we consider the case of PZT with $x=0.50$: if we divide a large 
domain (say, with dimensions of the order of 1 $\mu$m) to cubes containing $N$ primitive cells, roughly $2/3$ of 
the cubes contain $N_{\mathrm{Zr}}=xN$ Zr ions with $Np-\sqrt{Npq} \leq N_{\mathrm{Zr}}\leq Np+\sqrt{Npq}$. If we 
now take the cube edge to be 10 nm (typical spot sizes used in TEM and ED studies are between 1 and 5 nm), 
almost $1/3$ of the cubes have $x<0.49$ or $x>0.51$ (for large values of $N$ binomial distribution can 
well be approximated by a Gaussian distribution). Spatial composition variation is the probable origin explaining 
why some domains are rhombohedral, whereas some are monoclinic. In a transition region (compositions between well 
established rhombohedral and tetragonal regions) $Cm$ phase serves as a transition bridge. The idea of an inhomogeneous 
distribution of Zr and Ti ions is also consisted with the TEM and ED observations according to which there is a 
rather large spatial variation in ED patterns as discussed above. Surface phenomena are sensitive to local distortions. 
It is the feature of the binomial distribution that the variance is largest at around MPB composition ($p \approx q$) 
and is zero only for pure PbTiO$_3$ and PbZrO$_3$. This is seen from the line widths which increases with increasing 
$x$ when $0 \leq x \leq 0.50$. This is supported by an atomic pair distribution function analysis carried out for PZT 
powders with $x=0.40, 0.52$ and $0.60$ \cite{Dmowski}: the MPB is a crossover point with maximum disorder. 
It is also important to note that even if the average composition of different grains would be exactly the same, the spatial 
variation can correspond to domains with different symmetries. In practice, some composition variation in average 
compositions of different domains and grains exists, as the above mentioned solubility gap suggests. Also the composition of 
grain boundaries is expectably different from the inner parts. The most important factor behind the two-phase 'co-existences' 
is probably related to the differences in average compositions of different domains: the domains with larger Zr content can 
entirely transform to rhombohedral phase, while it looks that the phase transformation cannot be completed in those domains 
containing larger amount of Ti but is frozen to $Cm$ phase. We note that $Cm$ phase was interpreted to relieve the stress 
which otherwise would be generated due to the \emph{interacting} rhombohedral and tetragonal domains in ref. \onlinecite{Topolev}. 
From this point of view it is not a surprise that the Bragg reflections of the $Cm$ phase are much wider than those of the 
$R3c$ phase, and are dependent on temperature. It is our opinion that PZT with composition in the vicinity of the MPB (and 
similar systems) should be modelled as a two-phase system.

On the other hand, if one could prepare an ideally ordered PZT sample with $x=0.50$, $Cm$ phase might not be 
stable. To see this, we note that in the case of ordered PZT with $x=0.50$ the Zr/Ti distribution is 
homogeneous in the primitive cell scale and any deviation from this order results in less homogeneous 
distribution. Correspondingly, there are no Zr rich or deficient areas and in this sense also the mechanism 
favouring $Cm$ would be missing: if there are no local distortions due to the spatial composition variation, 
there likely are no domains with different symmetries (at least not so abundant as the experiments reveal), 
which in turn implies that there is no need for a phase serving as a bridge between rhombohedral and tetragonal 
phases. This might be important for a very small grain size powders where most of the grains would 
be consisted of one domain. Experimentally the two extrema cases, perfect disorder and order, can easily 
be distinguished by x-ray or neutron diffraction techniques. As the studies dedicated to double perovskites 
show, ordering is clearly seen from the superlattice reflections in x-ray (for instance, see ref. \onlinecite{Yukari1} 
where polymerized complex method was used to prepare Ba$_2$MnWO$_6$ double perovskites) and neutron powder 
diffraction patterns (see, e.g., ref. \onlinecite{Azad}). Also Raman spectra are consisted with the observed diffraction 
patterns \cite{Liegeois,Yukari2}. As the numerous studies show the difference in ionic radii (and, to a lesser 
extent, charge) does not provide sufficient driving force for a formation of the double perovskite structure. As 
proposed in ref. \onlinecite{Bell}, partial ordering may occur, although it is not easy to determine the degree of ordering 
in the present case. Even though it may not be possible to prepare double perovskite PZT samples the problem can 
be approached by computational techniques. Finite temperature 
Monte Carlo simulations carried out for PZT with a broad range of $x$ and random distribution of Zr and Ti in the 
$B$-cation site yielded a phase transition sequence $P4mm \rightarrow Cm \rightarrow R3m$ \cite{Vanderbilt} (no 
interaction between different domains was incorporated in this study). Virtual crystal approximation (VCA, where 
Zr and Ti ions are replaced by the same fictitious average atom) does not reveal a $Cm$ phase \cite{Vanderbilt}. 
It is an open question if VCA could reproduce $Cm$ phase when the interaction between domains is taken into account. 
After all, $Cm$ phase always co-exists with rhombohedral phase. The two-phase co-existence in the vicinity of the MPB has 
proven to be very puzzling and sometimes one can find reports according to which an 'ideal' sample preparation route could 
yield single phase samples. Recently, an equilibrium phase diagram satifying the Gibbs phase rule was given in 
ref. \onlinecite{Rossetti}. The essential result was that, by the Gibbs phase rule, the single-phase fields on the phase diagram 
must be separated by two-phase regions (and $Cm$ phase is by no means an exception)\cite{Rossetti}.  
According to ref. \onlinecite{Fu}, the flat energy surface near the rhombohedral phase may be the most important factor for 
high piezoelectric constant materials. As the co-existence suggests, this seems to be the case in PZT system.

\section{\label{ALB}Anisotropic line broadening.} 
The anisotropic ($hkl$-dependent) line broadening was ascribed to a 'microstrain' in ref. \onlinecite{FranttiJPCM}. 
There are several factors which together are responsible for the anisotropic line broadening: (i) Zr substitution for Ti creates 
local strains due to the size difference, (ii) spatial composition variation: Zr-rich clusters take larger 
volume per formula unit than Ti-rich clusters, and (iii) Pb-ion shifts depend on the $B$-cation environment (i.e., 
Pb-shifts vary spatially as a response to the spatial variation of the $B$-cations). Although the given mechanisms are 
interweaved, the given division serves as a way to 
estimate which one is dominating at different composition regimes. Thus, the changes in Bragg reflection intensities 
and displacement parameters in the case of small Zr/Ti substitutions in PbTiO$_3$/PbZrO$_3$ can largely be understood 
by considering the mechanism (i). Thus, following the treatment given in ref. \onlinecite{Zevin}, small Zr substitution 
for Ti in PbTiO$_3$ might be treated by considering the concentration of Zr atoms $x$, the difference between 
matrix (Ti) and solute (Zr) atoms $\Delta R$. Now, the degree of distortion is characterized by the mean square atomic 
displacement $\langle U^2 \rangle=jx(\Delta R^2)$, where $j$ is a constant. Static atomic displacements cause a 
reduction in diffraction peak intensity and increase in diffuse scattering and can be treated in the same way as 
thermal vibrations. However, this treatment is no more adequate for intermediate values of $x$ since there is no 
way of selecting 
such a matrix that experimentally observed features could be explained. Although the mechanism (i) can explain changes 
in intensities (matrix and solute atoms have different scattering lengths and cross sections) and diffuse scattering 
(seen as an increased intensity in the Bragg reflection tail regions: the peak width at half maximum is practically not 
affected by this mechanism) for small values of $x$, it does not provide a model for changes in Bragg reflection widths. 
This is why the phenomenological model given in ref. \onlinecite{Stephens} was adopted: it also takes into account the 
$hkl$-dependent Bragg reflections widths by introducing a microstrain broadening 
$\Gamma_S^2=\sum_{HKL}S_{HKL}h^Hk^Kl^L$, $H+K+L=4$. Laue symmetry imposes restrictions on the allowed $S_{HKL}$ terms. 
Expressions for $\Gamma_S^2$ for each Laue symmetry are given in ref. \onlinecite{GSAS}.

In addition, we note that rather strong intensity is commonly observed between the $h00$ 
and $00l$ with $h=l$ Bragg reflections, 
which was assigned to locally disordered regions in the vicinity of the domains walls in ref. \onlinecite{NohedaB}. 
Especially at room temperature it is practically impossible to assign a precise structural model for this phase, since 
it is not well crystallized. It might also be a precursor phase which at higher Zr contents evolves into a rhombohedral 
phase. This interpretation was given to the data collected on PHT powders, see Fig. 4 in ref. \onlinecite{FranttiSub}. 
Nevertheless, scattering from this phase adds intensity to certain diffraction peaks, which in practise also contributes 
to an asymmetric line broadening (as it is not possible to unambiguously separate it: it is the sum from different 
contributions what one observes in the case of powder diffraction). For example, the larger and smaller $d$-spacings 
sides of the pseudo-cubic $200$ and $002$\mbox{ }reflections, respectively, seem to gain intensity. Particular care is 
necessary once this phase is modelled and generally one might do better by collecting data on samples with different Zr 
contents and at different temperatures.

\begin{figure}
\begin{center}
\includegraphics[width=8.6cm]{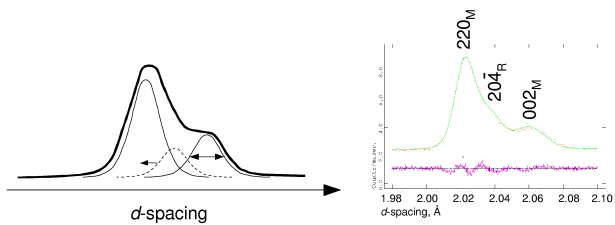}
\caption{\label{Asymmetry} Left panel: Schematic illustration of the 
way how the neglect of anisotropic line broadening affects the 
position of the rhombohedral peaks: once the rightmost peak broadens, 
it pushes the rhombohedral peak (middle) towards smaller $d$-spacing. 
Righ panel: Observed and computed high resolution neutron powder 
diffraction profile collected on PZT with $x=0.53$ at 4 K. Pseudo-cubic 
$200$\mbox{ }reflection region is shown. Note that the $002_M$ peak 
($Cm$ phase) at around 2.06 \AA\mbox{ }is significantly broader than 
the $220_M$ ($Cm$ phase) peak at around 2.02 \AA. Now, the position 
of the rhombohedral $20\bar{4}_R$ peak depends on the way the line 
broadening is taken into account. 
Figure adapted from ref. \onlinecite{FranttiArxiv}
}
\end{center}
\end{figure}

Although a reasonable structure refinements for medium resolution data 
collected on PZT samples with $x\leq 0.50$ were obtained using a 
lineshape which ignored anisotropic line broadening, the situation was 
quite different for high resolution facilities, particularly once the 
data were collected on two phase samples. In such 
a case it was essential to use an appropriate profile function. 
\emph{Now, when the anisotropic line broadening is neglected}, the 
fit in the case of certain weak peaks becomes slightly worse (as was 
observed to be the case of weak superlattice reflections, which are 
less weighted in the refinements). The reason for this is illustrated 
in Fig. \ref{Asymmetry}: 
in order to improve the fit corresponding to the strong pseudo-cubic 
$200$ reflections, during the refinement the position of $R3c$ 
reflection is shifted toward higher $d$-spacings, which in turn 
resulted in a small shift of the weak peaks at 2.44 and 1.06 
\AA. The anisotropic line broadening was not limited to the $d$-spacing at around 2 \AA, 
and similar mechanism was seen at other $d$-spacings.  A structural model which 
reduces the symmetries to 'model' the $hkl$-dependent line broadening takes local
 disorder into account incorrectly, since in this way the disorder is 
assigned to be periodical, obeying the space group symmetry. 

\section{\label{TPM}Two-phase models}
\subsection{Pb(Zr$_{0.52}$Ti$_{0.48}$)O$_3$ and Pb(Zr$_{0.53}$Ti$_{0.47}$)O$_3$ samples}
\begin{figure}
\begin{center}
\includegraphics[width=8.6cm]{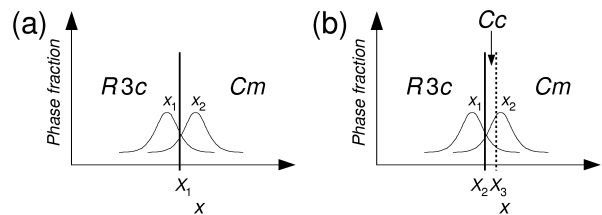}
\caption{\label{CompVar} The consequences of the spatial composition 
variation (at fixed temperature and pressure) in the case of two samples 
with average compositions $x_1$ and $x_2$ in the vicinity of (a) the phase 
boundary separating $R3c$ and $Cm$ phases (at $X_1$) and (b) two phase 
boundaries separating $R3c$ and $Cc$ (at $X_2$) and $Cc$ and $Cm$ phases 
(at $X_3$). 
Figure adapted from ref. \onlinecite{FranttiArxiv}
}
\end{center}
\end{figure}

In the context of low-temperature phases the role of spatial composition variation was 
discussed in refs. \onlinecite{FranttiPRB} and \onlinecite{FranttiJPCM}. 
The room- and low-temperature crystal symmetry of Zr rich PZT ceramics is $R3c$ 
\cite{Michel,Corker,FranttiJPCM}, except for the compositions with 
$x \approx 1$\cite{Comment3}. The room temperature symmetry of PbZrO$_3$ is $Pbam$ \cite{Glazer2,Corker2}.
Since spatial composition variation cannot be completely eliminated, there 
must exist two phases in the vicinity of MPB. Fig. \ref{CompVar} (a) illustrates the consequences 
of spatial composition variation in the vicinity of MPB at low temperature. 
To allow the existence of $Cc$ phase necessitates that there should be a narrow region in the $x-T$ plane 
were this phase is stable or metastable. Now, if we were to explain the existence of $Cc$ phase, two phase boundaries 
located somewhere between $0.52 \leq x \leq 0.54$ should be assumed to exists, 
see Fig. \ref{CompVar} (b). This in turn, once the spatial composition variation is taken into account, 
leads to three phase 'co-existence'. The simplest $Cm+R3c$ model (corresponding to Fig. \ref{CompVar} (a)) was 
preferred in refs. \onlinecite{FranttiPRB} and \onlinecite{FranttiJPCM}, since it was able to 
explain the experimental observations in simplest terms. At low temperature 
(4 K\cite{FranttiJPCM} or 10 K \cite{FranttiPRB}) the phase fraction of 
the $Cm$ phase was monotonically decreasing with $x$ increasing from $0.52$ 
to $0.54$, which in turn implies that two-phase 'co-existence' is 
predominantly due to the spatial composition variation\cite{Comment4}. 

It is worth to point out that the model proposed in ref. \onlinecite{Ranjan} was ruled out also in a recent paper, see 
the last paragraph in ref. \onlinecite{Cox}. After all, the only difference between  $Cm+R3c$ model and $Cm+Cc$ model 
proposed in ref. \onlinecite{Cox} is that $R3c$ symmetry was used (corresponding to two lattice parameters and four 
atomic coordinates) instead of the $Cc$ symmetry (corresponding to four lattice parameters and twelve atomic 
coordinates, although constraints were used to decrease refinable parameters in ref. \onlinecite{Cox}). Now, $Cc$ is a 
subgroup of $R3c$ space group. Since no superlattice reflections characteristic only to $Cc$ phase were observed 
(such should appear at large $d$-spacing region, if symmetry is lowered from $R3c$ to $Cc$), $R3c$ symmetry was 
favoured in ref. \onlinecite{FranttiPRB}. This allowed considerably simpler structural model: one is tempted to ask how 
much can be gained by first decreasing the symmetry from $R3c$ to $Cc$ and then decreasing the number of refinable 
parameters by introducing constraints. This issue is discussed on a more general ground in ref. \onlinecite{Massa}. 
In ref. \onlinecite{Cox} the use of $Cc$ phase was largely based on the TEM and ED observations \cite{Noheda2}
according to which superlattice reflections, which were inconsistent with $R3c$ symmetry, were observed at low temperature. 
This observation is consistent with the observations reported in ref. \onlinecite{Glazer} but the interpretation is entirely 
different. Now, it would be interesting to consider: (i) similar possibilities as were given in ref. \onlinecite{Glazer} 
(summarized above), and (ii) if the observed features are only characteristic to TEM samples (layers with thickness 
between 5 and 100 nm were studied and it was further stated that the size distribution of $Cc$ phase ranged 
between 3 to 10 nm\cite{Noheda2}. This is contrasted by the observation according to which the line widths of $R3c$ phase 
were much narrower than those of $Cm$ phase\cite{FranttiJPCM}. Neither of the phases was in a form of 'nanoparticles'. 
It would also be interesting to see how large octahedral tilts are necessary to 'reproduce' the ED patterns, including 
those superlattice reflections which were taken as an evidence for $Cc$ symmetry, since they were not observed by NPD studies.
It is an often overlooked issue that $Cc+Cm$ and $Cm+Cc$ models, given in refs. 
\onlinecite{Ranjan} and \onlinecite{Cox} respectively, 
are \emph{completely different} (and mutually exclusive): they present models 
where $Cm$ and $Cc$ symmetries were swapped. Although this swapping might not be 
so obvious from studies solely based on TEM/ED techniques it is clearly seen from 
X-ray powder diffraction and NPD models. In the case of TEM/ED techniques very small 
volumes are studied in a time (this can also give wrong idea about the phase purity, 
i.e., if the sample with macroscopic dimensions contains one or two phases) and often 
no attempts to determine sufficiently accurate lattice parameter were done.
Whatever is the reason for overlooking this issue, this type of space group swapping 
results in entirely different structural parameters and phase fractions. Needless to 
say, this has nothing to do with composition variation or differences in crystallite 
sizes, see also related discussion in ref. \onlinecite{NohedaArxiv}. Despite the 
increased number of refined parameters, the residuals were still rather high and the 
differences between different models compared in ref. \onlinecite{Ranjan} were marginal. 
For instance, $R_{exp}$ was lowest for the $Cm+R3c$ model which was rejected in ref. 
\onlinecite{Ranjan}. As discussed in the next section, the model given in ref. 
\onlinecite{Ranjan} does not assign octahedral tilts to a correct phase. In constrast, the models 
proposed in refs. \onlinecite{FranttiPRB} and \onlinecite{Cox} assigned octahedral 
tilts to the same phase. In the case of the data shown in ref. \onlinecite{Ranjan} 
the intensity of the peak(s) at around $1.06$ \AA\mbox{ }was barely above the 
noise level and the $Cc+Cm$ model assigns more reflections than there are data 
points in this region. Thus these 
reflections do not provide justification for the use of $Cc$ phase.

In addition, some reports claimed that dielectric measurements gave support to $Cc$ phase (for example, see ref. \onlinecite{Ranjan3}). 
Although these types of measurements might be useful once phase transitions in single phase samples are studied, not too much emphasis should 
be put on data collected on multiphase samples. In this context we note that the samples studied in refs. \onlinecite{Ranjan2} and \onlinecite{Ragini3} 
contained a nonidentified impurity phase(s), as was revealed by the peaks at around 28 and 35 two-theta degrees (corresponding to $d$ spacings 
3.18 \AA\mbox{  }and 2.56 \AA, respectively), in addition to the aforementioned two-phase co-existence. This implies that the composition 
was not well known and thus the possibility of compositional and structural inhomogeneities should be kept in mind. 
Although this non-perovskite phase was well resolved, it was not included in the Rietveld refinement model considered in ref. \onlinecite{Ranjan2}, 
which in turn results in an error in the structural parameters of the perovskite phase(s). This is likely related to the fact that the 
diffraction pattern shown in ref. \onlinecite{Ranjan2} (reported to have $x=0.52$) is reminiscent to the diffraction pattern with $x=0.53$ shown 
in ref. \onlinecite{FranttiPRB}. In contrast, the diffraction pattern with $x=0.52$ given in ref. \onlinecite{FranttiJJAP}) and the diffraction pattern 
of the $x=0.52$ sample reported in refs. \onlinecite{Cox} and \onlinecite{NohedaA,NohedaB} were reminiscent. 

\subsection{\label{JPCM} Pb(Zr$_{0.54}$Ti$_{0.46}$)O$_3$ sample}

The anisotropic line broadening and the absence of the superlattice reflection evidencing $Cc$ symmetry in PZT powders 
with $x=0.52$ and $x=0.53$ provided a motivation to carry out a subsequent study using a high resolution NPD instrument\cite{FranttiJPCM}. 
To model the peak profiles, GSAS\cite{GSAS} lineshape 4 by Stephens \cite{Stephens} was used in this study.  In addition, Zr content was 
slightly increased to $x=0.54$\cite{FranttiJPCM} (although it was possible to fit \emph{all} reflections using the $Cm+R3c$ model also 
in the case of $x=0.52$ and $x=0.53$\mbox{ }samples). This allowed a more reliable refinement and symmetry identification to be done by 
studying the \emph{changes in phase fractions versus temperature}. Importantly, for these compositions it has been found that oxygen 
octahedra tilts increase with increasing $x$ (see ref. \onlinecite{Corker}) and decreasing temperature (this feature is discussed in refs. 
\onlinecite{Thomas} and \onlinecite{FranttiPRB}). We also note that previously $Cc$ was proposed to be a space group symmetry corresponding to high 
isotropic pressure\cite{Haines}, whereas the present review concentrates on the determination of space group symmetries versus composition 
and temperature at ambient pressure. Now, PZT sample with $x=0.54$ provided a test for clarifying which phase is the preferred one at low 
temperature. This is of particular interest also from the point of view of \emph{ab initio} computations dedicated for PZT according to 
which the largest piezoelectric $d_{33}$ coefficients are found to be large namely in the rhombohedral side of the MPB: much smaller values 
were found in the tetragonal side of the MPB \cite{Vanderbilt2}. These observations are consistent with the experimental observations on 
PZT\cite{Du} and Pb(Zn$_{1/3}$Nb$_{2/3}$)O$_3$-PbTiO$_3$ (PZN-PT) and Pb(Mg$_{1/3}$Nb$_{2/3}$)O$_3$-PbTiO$_3$ (PMN-PT)\cite{Park} (the two 
latter systems also possess MPB separating rhombohedral and tetragonal phases). The NPD data in ref. \onlinecite{FranttiJPCM} suggests the following 
phase transition sequence with decreasing temperature in the vicinity of MPB: 
$Pm\bar{3}m \rightarrow P4mm \rightarrow Cm \rightarrow R3m \rightarrow R3c$ where adjacent groups have a group-subgroup relationship.
Due to the composition variation some domains undergo the same phase transition sequence at higher (Zr richer) or lower 
(Ti richer) temperatures. This sequence demostrates the role of $Cm$ phase linking $P4mm$ and $R3m$ phases. The co-existence 
of the rhombohedral and $Cm$ phase is consistent with the idea that they are energetically almost as favourable and rather 
small changes in temperature, pressure or an applied electric field can transform $Cm$ phase to rhombohedral phase. 
First-principles computational study revealed that the electric field induced phase transformation sequence with an 
increasing applied electric field was found to be reversed when the field was released \cite{Vanderbilt2}, whereas the 
same was not found experimentally in ref. \onlinecite{Noheda3}. This different behaviour was attributed to defects\cite{Vanderbilt}: 
the model used in \emph{ab initio} computations was defect free, whereas in practice they cannot be completely eliminated. 
In this context it is interesting to note the domain switching in Pb(Zr$_{0.49}$Ti$_{0.51}$)O$_3$ sample, where tetragonal and rhombohedral 
(volumetric phase fractions 79 and 21 \%, respectively) phases were present, studied through in situ uniaxial compression experiments \cite{Rogan}. 
Data were analysed via Rietveld refinement which allowed texture and lattice strains to be simultaneously determined. It was found that the 
rhombohedral phase responds more readily both to electric fields and applied stress. 

As is seen from Fig. 1 in ref. \onlinecite{FranttiJPCM}, once the crystals are compressed with 
decreasing temperature, the $R3c$ phase was favoured and $Cm$ phase did not transform to $Cc$ 
phase, in strong contrast with the model proposed in ref. \onlinecite{Ranjan}). We note that the 
volume of the $R3c$ phase grows with decreasing temperature (for instance, see Fig. 1 in ref. 
\onlinecite{FranttiJPCM}). Thus, no evidence was found for the $P4mm \rightarrow Cm \rightarrow Cc$ 
phase transition sequence. When the oxygen octahedra had almost no chance to contract, they were 
rotated to opposite directions along the pseudo-cubic $111$ axis (tilt system $a^-a^-a^-$). This 
tilt results in a decrease of the volume of cuboctahedra around Pb ions. Changes evidencing  
$Cc$ symmetry (and the corresponding tilt system) were not observed. \emph{Indeed, the phase 
fraction of $Cm$ phase (which was assigned to $Cc$ symmetry in ref. \onlinecite{Ranjan}) decreased with 
decreasing temperature, whereas $R3c$ phase fraction significantly increased with decreasing 
temperature.} This can be confirmed by a naked eye observation of the diffraction patterns in Fig. 1 
in ref. \onlinecite{FranttiJPCM}, although the results of refinements were also given. The intensity of 
the peak at around 1.06 \AA\mbox{  }was increasing with decreasing temperature, which further 
confirmed that its origin is the phase assigned to $R3c$ phase. Also the peak at around 1.06 
\AA\mbox{  }was well fit by $R3c$ symmetry. In other words, in ref. \onlinecite{Ranjan} $Cc$ phase was 
used to model the Bragg reflection from the pseudo-tetragonal phase and also the superlattice 
reflections. Thus, $Cc$ symmetry was used to model Bragg peak from two distinct phases, which is 
not correct. However, the temperature and composition dependence clearly show that when the phase 
fraction of $Cm$ phase decreased, the intensity of the superlattice reflection was increasing 
together with the phase fraction of $R3c$ phase. This behaviour is consistent with the composition 
variation: the distortion (monoclinic or rhombohedral) and the volume of the distortion depend on 
average composition and temperature.

\section{Correlation between the $c/a$ axis ratio and piezoelectric constants}

\begin{figure}
\begin{center}
\includegraphics[width=6.6cm]{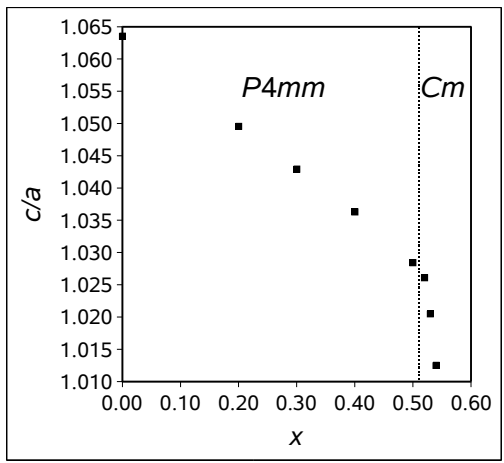}
\caption{\label{AxisRatio} Experimentally determined room temperature $c/a$ axis values for PZT ceramics. 
Lattice parameters were adapted from ref. \onlinecite{GlazerMabud} (PbTiO$_3$) and refs. \onlinecite{FranttiPRB} and 
\onlinecite{FranttiJPCM} (PZT).}
\end{center}
\end{figure}

It is of interest to compare the recent computational results with the structural parameters obtained through structural studies. 
According to the computations carried out for PbTiO$_3$ and Pb(Zr$_{0.50}$Ti$_{0.50}$)O$_3$ (where Zr and Ti atoms were alternatively 
occupying the $B$-cation layers perpendicular to the $c$ axis) the fixed volume piezoelectric constants $e_{33,v}=e_{33}-0.5(e_{31}+e_{32})$ 
and $e_{13,v}=e_{13}-0.5(e_{11}+e_{12})$ become very large in the range $1.0 <c/a <1.02$ \cite{Wu}. Experiments have shown that the $c$ axis 
remains almost constant for $0 < x < 0.50$, whereas $a$ axis increases almost linearly for these compositions, corresponding to a steady decrease 
of the $c/a$ axis ratio \cite{FranttiPRB}. However, in the MPB region the $c/a$ axis ratio decreases dramatically: averaging monoclinic $a$ and $b$ 
axes the $c/a$ is found to be $1.01$ for $x=0.54$ \cite{FranttiJPCM}. Fig. \ref{AxisRatio} shows the room temperature $c/a$ axis ratios. It is seen 
that the ratio strongly decreases once $x$ increases from $0.50$ to $0.54$. This is consistent with the idea that $Cm$ phase continuously transforms 
to rhombohedral phase. For pure PbTiO$_3$ \emph{ab initio} computations show that $P4mm$ phase is energetically more favourable than the $R3m$ phase, 
consistently with the experimental observation. This means that the electric field, which in practice can be applied (without electric breakdown), 
does not induce a phase transition 
sequence $P4mm\rightarrow Cm \rightarrow R3m$ and, via the polarization rotation model, piezoresponse remains much lower than in the case of PZT in 
the vicinity of the MPB. However, recent computational study demonstrated that pressure can induce a MPB in PbTiO$_3$ \cite{WuCohen}. It was found that 
pressure induced a phase transition sequence $P4mm \rightarrow Cm \rightarrow R3m \rightarrow Pm\bar{3}m$, together with large piezoelectric coupling 
coefficients in the transition regions\cite{WuCohen}.

\section{\label{BVS}Bond-valence sums and ion valences}
Computations of bond valence sums (BVS) provides a way to test whether the bond lengths are reasonable. 
Although it is not sensitive for minor changes in symmetry, unless bond lengths are altered, it has proven 
to be useful for confirming the effect of the proposed Pb, Zr and Ti ion displacements\cite{FranttiPRB}. 
Namely, if Pb ions are constrained to be located in their ideal positions in the case of $P4mm$ phase, 
the Pb ion valence would be $\approx +1.8$, whereas it is quite close to the nominal valence $+2$ once 
Pb ions are allowed to be displaced towards $\langle 110 \rangle$ directions. This is actually rather general tendency for Pb 
ion and occurs in rhombohedral phases. This is quite plausible since the cuboctahedra is so large that if Pb 
ion would occupy its centre, it would result in valence deficient Pb ions. Thus, by forming four short bonds with 
oxygen more reasonable valence values for Pb are achieved. Similarly, the average $B$-cation valence 
$v(B_{av})=xv(\mathrm{Zr})+(1-x)v(\mathrm{Ti})$ was $+4$ when Zr and Ti ions were allowed to have different 
fractional coordinates. In the case of the tetragonal PZT it was demonstrated that if one insists to 
constrain the fractional coordinates to be the same, no Wyckoff $1b$ position in the unit cell 
would correspond to the nominal valence $+4$. It also worth to point out that if Pb ions are constrained to 
be at the $1a$ site symmetry position Pb ADPs are very large. On the other 
hand, by constraining the $B$ cations to have the same fractional coordinates yielded negative ADP values. 
Allowing Pb ions to be displaced towards $\langle 110 \rangle$ directions and Zr and Ti ions to have different 
fractional $z$-coordinates resulted in physically reasonable ADP values. 
From this point of view it was important to check the validity of the structural models by confirming that 
both anomalous valence and ADP values can be eliminated by the same structural model. Oxygen valences in 
PZT systems were close to their nominal valence value $-2$. Also oxygen ADP values were reasonable. Thus, as 
far as NPD studies are considered, it is a decent approximation to assume oxygen octahedra to fullfil the 
requirements of space group symmetry, in contrast to the case of cation positions. This also implied that it was 
not necessary to invoke the plausible mechanism, the expansion of Zr octahedra and contraction of Ti octahedra, 
to correct the anomalously large Zr and small Ti valences in the structural models used to model the NPD data.

BVS were found to be consistent with the observed phase transition sequences so that reasonable valence 
values for each ion were obtained also in the case of the $Cm$ and $R3c$ phases. The relationship between 
the oxygen octahedra and cuboctahedra volumes and octahedra tilts were pointed out in ref. \onlinecite{Thomas}. 
Namely, $R3c$ symmetry allows oxygen octahedra tilt, which means that the volume of oxygen octahedra increases 
with increasing tilt angle. This in turn allows the valence of each ion to be close to their nominal values. 
The same is not possible in the case of $R3m$ symmetry and thus the observed oxygen octahedra tilts versus 
temperature and $x$ seen in the case of PZT samples with $x=0.52$ and $x=0.53$ are consistent with the 
'constraints' set by nominal valences\cite{FranttiPRB}.
In many ways similar findings, based on density functional theory computations carried on PZT modelled with large supercells, 
were reported in ref. \onlinecite{Grinberg}. Perhaps the most essential difference between the typical Rietveld refinement models and 
supercell models for PZTs is that the space group symmetries used in the latter approach allow different size oxygen octahedra 
for Zr and Ti, whereas they are constrained to be the same in Rietveld refinements (one actually replaces both Zr and Ti by a 
'pseudo-atom' whose scattering length (NPD) or cross section (X-ray diffraction) is a weighted average over Zr and Ti ions). This 
explains why the individual bond-valences for Zr and Ti were found to be close to their nominal values in the case of the supercell 
computations \cite{Grinberg}. 
In this sense it is surprising to see that both approaches support the idea that oxygen position do not suffer from local 
distortions to the extent cation positions do. 

\section{Conclusions}
The pitfalls related to the modelling of the crystal structures of PZT ceramics with different compositions and at different 
temperatures were reviewed. The role of the spatial composition variation, resulting in two-phase 'co-existence' in the vicinity 
of the phase boundaries and anisotropic line broadening, were reviewed. The conditions set to sample preparation, powder diffraction 
instrument and line shape were addressed. Focus was put on the structure analyses on PZT ceramics with compositions in the vicinity 
of the morphotropic phase boundary (MPB). Recent studies pointed out that the two-phase co-existence in the vicinity of the phase 
boundary is a thermodynamical necessity in PZT system and further is very crucial for the high electromechanical coupling coefficients 
observed in MPB compositions. Two different approaches for modelling the two-phase structures of PZT ceramics with compositions 
corresponding to the MPB were identified. The first is a method were space group symmetry was decreased so that both local and average 
crystal symmetries were modelled, whereas the second method was based on the highest space group symmetry compatible with the 
observed Bragg reflections and where local distortions were taken into account by selecting an appropriate line shape. Recently 
published $Cm+Cc$ and $Cc+Cm$ structure models were compared with $Cm+R3c$ model where anisotropic line shapes were used. 
Whereas $Cm+R3c$ model is a higher symmetry version of $Cm+Cc$ model, $Cc+Cm$ model was shown to be inapplicable. Once the local 
distortions, resulting in anisotropic line broadening, are correctly taken into account, we do not find support for $Cm+Cc$ model. 
Instead we found that $Cm+R3c$ model describes all Bragg peaks and their intensities well. It was further shown that the $Cm+R3c$ 
model is consistent with the structural features observed at other compositions and temperatures, which is particularly important 
in the vicinity of the phase boundary. Also issues related to the effects of a finite sample size used in transmission electron 
microscopy and electron diffraction (ED) studies were discussed and it was pointed out that while locally ordered areas may occur in 
nanosize samples, resulting in additional reflections in ED patterns, they frequently do not appear in bulk samples. Thus, the same 
structural model might not be valid for nanoscale and bulk samples. It was further pointed out that these features are not limited to 
MPB compositions. Spatial variation of Pb ions shifts follows from the spatial variation in the $B$-cation distribution. This is 
believed to be important for explaining the experimental structural data and for understanding electrical conductivity and 
piezoelectric properties. Connection between the structural parameters and piezoelectric properties were discussed in light 
of the recent computational studies.

\section*{Acknowledgments}
Author is grateful for the Academy of Finland for financial support (Project numbers 207071 and 
207501).


\begin{thebibliography}{99}
\bibitem{Comment3} The situation might be different for a high pressure case or at low temperatures in the 
case of the Zr rich PZT (e.g., for $x\geq 0.60$) since the octahedral tilting allowed by $R3c$ symmetry might 
no more be sufficient. If so, one could expect that this phase extends to a larger values of $x$. One may also 
speculate if symmetries allowing different octahedral tilts could be stabilized in the case of strained thin films.
\bibitem{Comment4} Although spatial composition variation is mainly responsible for the two-phase 
co-existence and line broadening phenomena, other factors, such as crystal size, have an 
effect on line broadening and probably affect phase fraction estimations.
\end{thebibliography}
\end{document}